\documentclass{article}
\usepackage[utf8]{inputenc}
\usepackage{authblk}
\usepackage{graphicx}
\usepackage{amsmath,amsfonts,amssymb}
\usepackage{hyperref}
\usepackage[square,sort,comma,numbers]{natbib}

\newtheorem{example}{Example}
\newtheorem{definition}{Definition}
\newtheorem{property}{Property}
\newtheorem{proof}{Proof}

\title{
    Inference of Common Multidimensional Equally-Distributed Attributes
}

\author[1,2]{Alejandro Álvarez-Ayllón\thanks{\href{mailto:alejandro.alvarez@uca.es}{alejandro.alvarez@uca.es}}}
\author[1]{Manuel Palomo-Duarte}
\author[1]{Juan-Manuel Dodero}

\affil[1]{
    Department of Computer Science and Engineering,
    University of Cadiz, Spain
}
\affil[2]{
    Department of Astronomy,
    University of Geneva, Switzerland
}

\newcommand{\eqdist}{\stackrel{d}{=}}
\newcommand{\X}{$X$}
\newcommand{\Y}{$Y$}
\newcommand{\RR}{$R$}
\newcommand{\RS}{$S$}
\newcommand{\AttrA}{$a_1,a_2,\dots,a_n$}
\newcommand{\AttrB}{$b_1,b_2,\dots,b_m$}

\begin{document}

\maketitle

\begin{abstract}
Given two relations containing multiple measurements – possibly with
uncertainties – our objective is to find which sets of attributes from
the first have a corresponding set on the second, using exclusively a
sample of the data. This approach could be used even when the associated
metadata is damaged, missing or incomplete, or when the volume is too big for
exact methods.
This problem is similar to the search of Inclusion Dependencies (IND),
a type of rule over two relations asserting that for a set of attributes
\X\ from the first, every combination of values appears on a set \Y\ from
the second.  Existing IND can be  found exploiting the existence of
a partial order relation called \emph{specialization}.
However, this relation is based on set theory, requiring the values
to be directly comparable. Statistical tests are an intuitive possible
replacement, but it has not been studied how  would they affect the
underlying assumptions. In this paper we formally review the effect
that a statistical approach has over the inference rules applied to IND
discovery.
Our results confirm the intuitive thought that statistical tests can
be used, but not in a directly equivalent manner. We provide a workable
alternative based on a ``hierarchy of null hypotheses'', allowing for the
automatic discovery of multi-dimensional equally distributed sets of
attributes.
\end{abstract}


\section{Introduction}
\label{sec:introduction}

Imagine an astronomer facing several data files containing raw astronomical
measurements, with little or no explanation about their schema. These files may
come from different surveys or different sets of observations, and the user can
only make the following educated guesses:

\begin{itemize}
    \item The populations are likely the same, or at least very similar (i.e. stars)
    \item A subset of the attributes is shared between the relations (i.e. brightness
        on different electromagnetic bands)
    \item This measurement has an associated uncertainty\cite{Stonebraker2009},
    either explicitly stated or not
    (i.e. random errors, instrument precision, floating point 
    precision \cite{dawson2008comparing})
\end{itemize}

The first intuition would be to run some kind of statistical test between all possible
pairs of columns, as the Kolmogorov-Smirnov\cite{Hodges1958} or
Wilcoxon\cite{Wilcoxon1945} tests. And this is likely a good starting point,
but we are left only with a set of pairwise correspondences that
may not be enough to cross-match tuples between files.

\begin{figure}[ht]
    \centering
    \includegraphics[width=\linewidth]{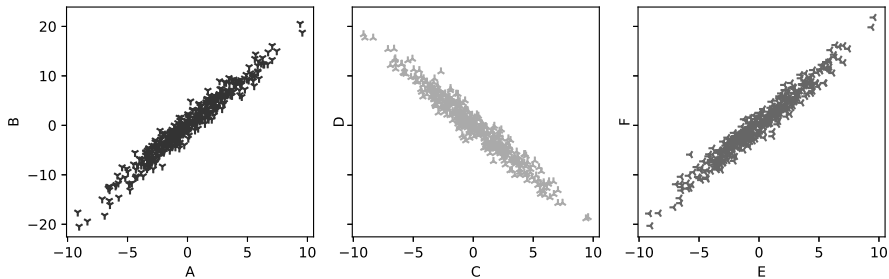}
    \caption{
        Example of a 2D distribution where the pairwise matching would not be
        accurate enough. It is artificial, but it serves to illustrate
        the point.
    }
    \label{fig:pairwise_ind}
\end{figure}

\begin{example}
In figure \ref{fig:pairwise_ind}, imagine that A and B are attributes from a relation
\RR, and C to E attributes from a relation \RS. Pairwise tests would tell
us that A matches C and E; and that B matches D and F. This information is
evidently not enough to do a cross-match.
\end{example}

Starting with this initial set of one-dimensional matches, one can pick
all the possible combinations of two attributes to find the potential
two-dimensional spaces where cross-matching could be attempted, and perform
another series of statistical multivariate tests to check for ``matching''
pairs (denoted as $\approx$).

\begin{example}
    Following our example, we could test after if $A,B \approx C,D$ and
    $A,B \approx E,F$.
\end{example}

At this stage, we would have $\binom{n}{2}$ possible options, $n$ being the number
of positive one-dimensional matches. In general, to look
for k-dimensional matching spaces we would have to test all possible
$\binom{n}{k}$ permutations, for any $k \le n$.

Unfortunately, this can quickly grow out of hand, with a combinatorial explosion
on the number of tests required at each increase of dimensionality. This becomes
impractical, in terms of computational run-time, even for a relatively
small number of attributes.
Furthermore, many of these tests will be redundant:
if we already know that $A, B,C \approx D, E, F$, it would seem that it does not make much
sense to test, say, $A, B \approx D, E$.

To complicate things even more, since we are performing statistical tests there
is always a possibility (bound by the significance level $\alpha$) of falsely
rejecting the equality of distribution. For instance, $A, B,C \approx D, E, F$ might
still be true even if $A, B \approx D, E$ is rejected.

The issue of finding higher dimensions where two pairs of set of attributes
still follow the same distribution resembles that of finding high
arity Inclusion Dependencies (IND) between two relational datasets.
On the other hand, uncertainties and statistical errors make the problem
different enough as to require a more careful consideration of their
effects on the foundations of IND finding algorithms.

In this paper, we discuss how to map the IND inference rules into
the problem of \emph{finding multidimensional equally-distributed set
of attributes}, and the limitations arising from the approximate nature of
statistical tests.

The rest of the paper is structured as follows: next, in section \ref{sec:background},
we introduce the background for the research. Then, in section \ref{sec:proofs}
we develop the proofs of the inference rules for numerical data.
Next, in section \ref{sec:discussion} we discuss the findings obtained an their
implications. Finally, in section \ref{sec:conclusion} we compile the conclusions of
the paper and the future work.

\section{Background}
\label{sec:background}

Let \RR\ and \RS\ be two relations, and \AttrA\ and
\linebreak \AttrB\ two sets of $n$ and $m$ attributes from both relations
respectively.

\begin{definition}
A rule of the form $\sigma = R[a_{i1},\dots,a_{ik}] \subseteq S[b_{i1},\dots,b_{ik}]$
(where $a_{i1},\dots,a_{ik}$ and $b_{i1},\dots,b_{ik}$ are
projections of \AttrA\ and \AttrB\ respectively)
is an \textbf{Inclusion Dependency} (IND) of \textbf{arity} $k \le min(n,m)$
between \RR\ and \RS. 
The particular case where $k = 1$ is also called an \textbf{Unary Inclusion Dependency} (uIND)\cite{Casanova1984}.
\label{def:ind}
\end{definition}

Note that definition \ref{def:ind} applies over the \emph{domains} of the
attributes, i.e. a potentially unlimited set of tuples where every possible
value from \RR\ and \RS\ is present.

Let $d$ be a concrete database instance from a database scheme $D$,
with finite samples from both relations.

\begin{definition}
An IND of the form $R[X] \subseteq S[Y]$ is \textbf{satisfied} (or valid) in $d$ 
if every combination of values from \X\ appears in \Y. This is denoted as $d \models \sigma$, where
$\sigma = R[X] \subseteq S[Y]$.
\label{def:satisfied}
\end{definition}

There are three inference rules that can be used to derive some
additional INDs from an already known set of IND\cite{Casanova1984}:

\begin{description}
    \item[Reflexivity] \hfill \\
        $R[X] \subseteq R[X]$
    \item[Permutation and projection] \hfill \\
        If $R[A_1,\dots,A_n] \subseteq S[B_1,\dots,B_n]$ then
        $R[A_{i_1},\dots,A_{i_m}] \subseteq S[B_{i_1},\dots,B_{i_m}]$ for each sequence
        $i_1,\dots,i_m$ of distinct integers from $\{1,\dots,n\}$
    \item[Transitivity] \hfill \\
        $ R[X] \subseteq S[Y] \land S[Y] \subseteq T[Z] \implies R[X] \subseteq T[Z]$
\end{description}

The second axiom is particularly important, as it can be applied
to derive a partial order relation which gives direction to the search space
of all possible IND\cite{DeMarchi2002}.
Let $I_1 = R[X] \subseteq S[Y]$ and $I_2 = R'[X'] \subseteq S'[Y']$.

\begin{definition}
$I_1$ \textbf{specializes} \cite{DeMarchi2002} $I_2$ - denoted $I_1 \prec I_2$ - iff

\begin{enumerate}
    \item $R = R'$ and $S = S'$
    \item $X$ and $Y$ are sub-sequences of $X'$ and $Y'$ respectively
\end{enumerate}

Equivalently, we can also say that $I_2$ \emph{generalizes} $I_1$.
\label{def:specialization}
\end{definition}

\begin{example}
$(R[AB] \subseteq S[EF]) \prec (R[ABC] \subseteq S[EFG])$. However,
$R[AB] \subseteq S[DE]) \nprec (R[ACD] \subseteq S[DFG])$, as AB and DE are not
sub-sequences of ACD and DFG respectively.
\end{example}

An important property of \emph{specialization} can be inferred \cite{DeMarchi2002}:

\begin{property}
    \label{prop:spec}
    Given $I_1 \prec I_2$
    \begin{enumerate}
        \item If $d \models I_2$, then $d \models I_1$
        \item By transposition, if $d \not\models I_1$ then $d \not\models I_2$
    \end{enumerate}
\end{property}

This property allows to quickly purge the IND search
space \cite{DeMarchi2002,koeller2002integration}:

\begin{enumerate}
    \item If we find that $d \models I_2$, we can ignore all $I_i$ s.t. $I_i \prec I_2$, since they will be satisfied
    \item If we find that $d \not\models I_1$, we can ignore all $I_j$ s.t. $I_1 \prec I_j$, since they will \emph{not}
        be satisfied
\end{enumerate}

\begin{example}
If we know that $R[AB] \not\subseteq S[EF]$, then we know that $R[ABC] \not\subseteq S[EFG]$.
\end{example}

\section{Inference rules for uncertain numerical data}
\label{sec:proofs}
Let's go back to the use case from our data scientist. Definition
\ref{def:ind} works over the \emph{domain} of the attributes,
but this is problematic when finding dependencies between attributes
that have the same domain, but different distributions.

\begin{figure}[ht]
    \centering
    \includegraphics[width=0.6\linewidth]{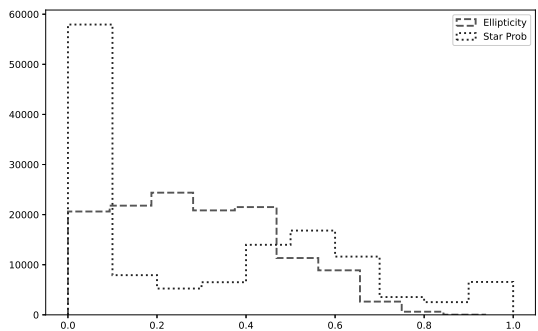}
    \caption{Two attributes with similar domains but significantly
    different distributions,
    from the Kilo-Degree Survey \citep{kuijken2019fourth}.}
    \label{fig:domains}
\end{figure}

\begin{example}
In a relation with galaxies and stars properties measured from images,
one may have for each tuple the aspect ratio of the ellipse that encompass
a given fraction of the light, and the probability of being a star.
Both are values from the domain $[0,1]$, but the distributions are nothing alike (Figure \ref{fig:domains}).
\end{example}

Thus, our user would likely be more interested in finding attributes that are \emph{identically distributed}:
$R[X] \eqdist S[Y]$.

\begin{definition}
Let $F_{X_1,\dots,X_k}$ be the cumulative distribution of the set of attributes
$X_1,\dots,X_k$, and $\mathbb{R}^k$ their domain.
$R[X_1,\dots,X_k] \eqdist S[Y_1,\dots,Y_k]$ (they are identically distributed)
if \cite{casella2002statistical}:

\begin{equation}
\begin{split}
F_{X_1,\dots,X_k}(x_1,\dots,x_k) = & F_{Y_1,\dots,Y_k}(x_1,\dots,x_k) \\
 & \forall (x_1,\dots,x_k) \in \mathbb{R}^k
\end{split}
\end{equation}
\label{def:eqdist}
\end{definition}

We could now replace the rule $\sigma$ in definition \ref{def:ind} by
$\sigma = R[X] \eqdist R[Y]$.

\medskip

\label{sec:caveat}

There is, however, one important caveat: this holds for the database
scheme $D$, where the true Cumulative Distribution Function (CDF) would be
defined. In our case, we are given a particular instance of the database,
which has a finite number of tuples. In other words, the database instance
could be seen as a \emph{sample} from an unknown database schema.

Consequently, we can only expect to use either the Empirical Cumulative Distribution
Function (ECDF), or a fitted curve (i.e. a Gaussian).
Either way, there will be uncertainty and definition \ref{def:eqdist} will not be directly
usable. Instead, it will be necessary to test for the null hypothesis
$H_0: P(R[X]) = P(S[Y])$, and this will be inherently affected by statistical errors
bound by the chosen significance level $\alpha$ and the power of the statistical test.

Nevertheless, we will show that the rules do apply assuming we know the
``true'' cumulative distribution. This will be at least enough to
guide the traversal of the search space, creating a ``hierarchy'' of
null hypotheses.

\medskip

Note that others have used statistical methods earlier to test IND,
but as an \emph{approximation} of the containment
rule \citep{Zhang2010, koeller2006heuristic}.
In our case, the question itself is statistical, so the inference rules
need to be re-evaluated.

\subsection{Reflexivity}
$R[X] \eqdist R[X]$

\begin{proof}
This property is trivial, as any random variable is distributed as itself.
\end{proof}

\subsection{Permutation and projection}
If $R[A_1,\dots,A_n] \eqdist S[B_1,\dots,B_n]$ then
$R[A_{i_1},\dots,A_{i_m}] \eqdist S[B_{i_1},\dots,B_{i_m}]$ for each sequence
$i_1,\dots,i_m$ of distinct integers from $\{1,\dots,n\}$.

\subsubsection{Permutation}

Considering that the CDF of $A$ could also be defined as
$P(a_1 \le A_1 \land \dots \land a_m \le A_n)$,
and that the logical operator $\land$ is commutative, it can be intuitive that the order
in which the attributes are specified does not affect their probability. However, we have
preferred to follow a different direction to prove that the relation $\eqdist$ is
invariant under permutation, since it is more general.

\begin{proof}
Let $f_{x_1,\dots,X_n} = \frac{\partial^n F_{X_1,\dots,X_n}(x_1,\dots,x_n)}{\partial x_1 \dots \partial x_n}$ be a joint probability \emph{density} function.
Let $(X'_1,\dots,X'_n)$ be a transformation $g$ of $(X_1,\dots,X_n)$ 
such that \linebreak $X'_i = g_i(X_1,\dots,X_n)$.
In general, the joint density function of $X'$ can be defined as

\begin{equation}
    f_{X'_1,\dots,X'_n}(X'_1,\dots,X'_n) = f_{X_1,\dots,X_n}(X_1,\dots,X_n) |J|
\end{equation}

Where $J$ is the Jacobian determinant of the inverse transformation $g^{-1}$ \citep{giri2014multivariate}.

In the particular case when the transformation
is defined by a non-singular matrix $M$ of size $n \times n$,
its Jacobian determinant is simply $|M|$, 
and the Jacobian determinant of the inverse transformation, $|M|^{-1}$ \citep{deemer1951jacobians},
which is a constant.
Given that the cumulative probability function is an integral over $f_{x_1,\dots,x_n}$, we can say that:

\begin{equation}
\begin{split}
    F_{X'}(X'_1,\dots,X'_n) & = F_{X}((X'_1,\dots,X'_n)M^{-1})|\det{M^{-1}}| \\
    & = F_X(X_1,\dots,X_n)|\det{M^{-1}}|
\end{split}
\label{eq:transform_cummulative}
\end{equation}

Let $X' = {x'_1,..,x'_n}$ and $Y'={y'_1,..,y'_n}$ be the result of a linear
transformation $M: R^k \rightarrow R^k$ over $X = {x_1,\dots,x_n}$ and $Y = {y_1,\dots,y_n}$
respectively. From definition \ref{def:eqdist} and equation \ref{eq:transform_cummulative}:

\begin{equation}
\begin{split}
X \eqdist Y \implies & F_{X}(x_1,\dots,x_k) = F_{Y}(x_1,\dots,x_k)
    \; \forall (x_1,\dots,x_k) \in \mathbb{R}^k \\
    \implies & F_{X}(x_1,\dots,x_k)|\det{M^{-1}}| = F_{Y}(x_1,\dots,x_k)|\det{M^{-1}}|
    \; \forall (x_1,\dots,x_k) \in \mathbb{R}^k \\
    \implies & F_{X'}((x_1,\dots,x_n)M) = F_{Y'}((x_1,\dots,x_n)M)
    \; \forall (x_1,\dots,x_k) \in \mathbb{R}^k \\
    \implies & F_{X'}(x'_1,\dots,x'_n) = F_{Y'}(x'_1,\dots,x'_n)
    \; \forall (x'_1,\dots,x'_k) \in \mathbb{R}^k \\
    \implies & X' \eqdist Y'
\end{split}
\end{equation}

This is true for \emph{all} one-to-one linear transformations
$M: R^k \rightarrow R^k$, of which a permutation is just a concrete case where $M$ is a permutation matrix.
\end{proof}


\subsubsection{Projection}

\begin{proof}
For the projection, we need to prove that if two sets of random variables
$X_1,\dots,X_n$ and $Y_1,\dots,Y_n$ are equally distributed, so are
any of their possible sub-sequences.

Let $X'$ and $Y'$ be the sequences $X_1,\dots,X_m$ and $Y_1,\dots,Y_m$ with $m < n$.
Their corresponding  CDF are just the \emph{marginal} CDF:

\begin{equation}
\begin{split}
    F_{X_1,\dots,X_m}(x_1,\dots,x_m) = & F_{X_1,\dots,X_m,X_{m+1},X_n}(x_1,\dots,x_m,x_{m+1},\dots,x_n) \\
    F_{Y_1,\dots,Y_m}(y_1,\dots,y_m) =& F_{Y_1,\dots,Y_m,Y_{m+1},Y_n}(x_1,\dots,x_m,x_{m+1},\dots,x_n)\\
    & \forall (x_1,\dots,x_m) \in \mathbb{R}^m \textrm{ and } x_i \xrightarrow{} \infty \; \forall i > m
\end{split}
\end{equation}

By definition \ref{def:eqdist}, the right hand-side of both equations must be the same.
By transitivity, 

\begin{equation}
\begin{split}
    & F_{X_1,\dots,X_m}(x_1,\dots,x_m) = F_{Y_1,\dots,Y_m}(y_1,\dots,y_m) \\
    & \implies X_1,\dots,X_m \eqdist Y_1,\dots,Y_m
\end{split}
\end{equation}

\end{proof}

\subsection{Transitivity}

$ R[X] \eqdist S[Y] \land S[Y] \eqdist T[Z] \implies R[X] \eqdist T[Z]$

\begin{proof}

\begin{equation}
\begin{split}
X \eqdist Y \land Y \eqdist Z \implies & F_X(x_1,\dots,x_k) = F_Y(x_1,\dots,x_k) \land \\
                                       & F_Y(x_1,\dots,x_k) = F_Z(x_1,\dots,x_k) \\
                              \implies & F_X(x_1,\dots,x_k) = F_Z(x_1,\dots,x_k) \\
                              \implies & X \eqdist Z
\end{split}
\end{equation}
\end{proof}

\section{Discussion}
\label{sec:discussion}
We have shown that replacing $\sigma$ in definition \ref{def:ind} with
$\sigma = R[X] \eqdist S[Y]$
leaves us with a similar set of inference rules that can be applied to support
the specialization relation from definition \ref{def:specialization}.

However, as we have already mentioned in section \ref{sec:caveat}, these rules
work if we know the true distribution of both sets of attributes.
In many cases, as it could be in astrophysics, the content of the attributes
are purely empirical, and we will have to approximate the definition
\ref{def:eqdist} with a statistical test with the null hypothesis
$H_0: R[X] \eqdist S[Y]$.

Nonetheless, we can apply the rule of projection to create a 
``hierachy'' of null hypotheses based on the definition of specialization, but
we will need to reformulate the property applied for the inference of new IND:

\begin{property}
Let $I_i$ be an assertion that two sets of attributes are equally distributed, and
$H_{0_i}$ the null hypothesis used to test it. Let $I_1 \prec I_2$.

\begin{enumerate}
    \label{prop:prob_spec}
    \item Accepting $H_{0_2}$ implies accepting $H_{0_1}$\footnotemark
    \item Rejecting $H_{0_1}$ \textbf{does not} imply the rejection of $H_{0_2}$
\end{enumerate}

\footnotetext{This is an abuse of terminology. Technically
\emph{not rejecting} $H_{0_2}$ implies that we can not reject $H_{0_1}$.}

\end{property}

The second part of this property can be simply explained by the fact
that a statistical test may falsely reject $H_{0_1}$ with a probability
given by the significance level $\alpha$. This is markedly different from
property \ref{prop:spec}, but still informative.

\begin{example}
If we have two sets of 10 attributes that are equally distributed, we have
$\binom{10}{3} = 120$ projections (specializations) of 3 dimensions
that must be equally distributed as well.
If we have a significance level of $\alpha = 0.1$, the expected number of
falsely rejected 3-dimensional equalities is 12. This can be used to check if the
actual number of rejections match the expectation.
\end{example}

\bigskip

We could try to apply a similar reasoning to the transitivity rule, but this
would arguably not work for inferring new IND properties:

\textbf{Projection} reduces the information available, since we remove dimensions.
If we can not reject the ``high arity'' null hypothesis, we \emph{should not}
reject any ``lower arity'' since, after all, there is less information available to do so.

\textbf{Permutation} does not alter the information available.
If we can not reject the null hypothesis for one permutation, we \emph{should not} reject
the null hypothesis for exactly the same data after being shuffled.

For \textbf{transitivity}, however, the information available for each test is
different and, therefore, nothing can be assumed. $X$ and $Y$ may be close enough
to not be possible to tell them apart, and the same may happen to $Y$ and $Z$. However,
$X$ and $Z$ may be separate enough as to be able to differentiate and
reject that they are equally distributed.

\section{Conclusion}
\label{sec:conclusion}

Helping data scientist to match and explore heterogeneous datasets, 
even when their scheme is unknown or unfamiliar, is an active
and interesting area of research with multiple ramifications \cite{Idreos2015,Milo2020},
one of which is schema matching \cite{Alawini2014}. To the best of our knowledge,
there has been no detailed discussions on how this can be achieved on multidimensional
spaces when uncertainty is unavoidable.

In this paper we have proven that inferring multidimensional sets of
``equally distributed'' attributes is feasible using similar mechanisms to those
of finding Inclusion Dependencies (IND) between two relational datasets.
In particular, the \emph{specialization} relation from definition
\ref{def:specialization} can be applied to give directionality to the search space,
and the property \ref{prop:prob_spec} provides capabilities to traverse it, avoiding
expensive combinatorial solutions.

However, this property can not be directly applied as a drop-in replacement
of the original property \ref{prop:spec}, as rejecting a low dimensionality
inclusion \emph{should not} necessarily cause the rejection of a higher
dimensionality one \emph{specialized} by it. This has to be taken into
account when adapting, or devising new, algorithms.

\subsection{Future work}

With this knowledge we can now start evaluating the viability of adapting
existing concrete solutions for the IND search problem to a more specific objective:
given two numerical datasets, with uncertainties, and without using the
associated metadata, find which subsets of attributes are ``equally distributed''.

As a non-exhaustive set of possible applications, once these sets of
attributes are found, they could potentially be used to
cross-match the objects between the relations\cite{Budavri2008}; to adapt
the dataset schemes and use them as a single one; or to apply a known
label from one to the other without knowing \textit{a priori} which
attributes can be used to do so.

\subsection*{Financial disclosure}
This research was funded by Spanish
National Research Agency (AEI), through the project VISAIGLE
(TIN2017-85797-R) with ERDF funds.


\end{document}